\documentclass{amsproc}

\usepackage{amsmath}
\usepackage{amsthm}
\usepackage{amssymb}
\usepackage{caption}
\usepackage{graphicx}
\usepackage[english]{babel}
\usepackage{tikz}

\title{Perturbative terms of Kac-Moody-Eisenstein series}

\author[P. Fleig]{Philipp Fleig}
\address{P. Fleig: Max-Planck-Institut f\"{u}r Gravitationsphysik (Albert-Einstein-Institut)\\ 
Am M\"{u}h\-lenberg 1, DE-14476 Potsdam, Germany\hfill\break
Freie Universit\"at Berlin, Institut f\"ur Theoretische Physik\\
Arnimallee 14, 14195 Berlin, Germany}
\curraddr{}
\email{}
\thanks{}

\author[A. Kleinschmidt]{Axel Kleinschmidt}
\address{A. Kleinschmidt: Max-Planck-Institut f\"{u}r Gravitationsphysik (Albert-Einstein-Institut)\\
Am M\"{u}hlenberg 1, DE-14476 Potsdam, Germany\hfill\break
International Solvay Institutes\\ ULB-Campus Plaine, C.P. 231, BE-1050 Brussels,  Belgium}
\curraddr{}
\email{}
\thanks{}

\dedicatory{Based on a talk given by the second author at ``3Quantum: Algebra-Geometry-Infor\-ma\-tion''(Tallinn, July 2012) on the work contained in~\cite{Fleig:2012xa}.}

\begin{document}

\begin{abstract}
Supersymmetric theories of gravity can exhibit surprising hidden symmetries when considered on manifolds that include a torus. When the torus is of large dimension these symmetries can become infinite-dimensional and of Kac-Moody type. When taking quantum effects into account the symmetries become discrete and invariant functions under these symmetries should play an important role in quantum gravity.  The new results here concern surprising simplifications in the constant terms of very particular Eisenstein series on the these Kac-Moody groups. These are exactly the cases that are expected to arise in string theory.
\end{abstract}

\maketitle

\section{Introduction}

In string theory, discrete dualities have played a central role in the research of the last 15 years.  These dualities can relate string theories on different backgrounds and with different matter content and are commonly referred to as U-dualities~\cite{HullTownsend,ObersUDualMTheory}. Their existence has led to the claim that there is a single M-theory underlying all string theories~\cite{Witten:1995ex}. 

A particular manifestation of this idea is given by type II superstring theory compactified on a $(d-1)$-dimensional torus, down to $D=11-d$ space-time dimensions. At low energies, the complete effective theory is maximal supergravity in $D$ dimensions and possesses a continuous $E_{d(d)}(\mathbb{R})$ hidden symmetry group~\cite{Cremmer:1978ds,Julia:1980gr}. These are maximally split Lie groups that for $6\le d\le 8$ are exceptional and for the other values of $d$ are defined for our purposes in Table~\ref{dualitygroups}. We will write $E_d(\mathbb{R})$ instead of $E_{d(d)}(\mathbb{R})$ for ease of notation. For $d>8$, the groups are infinite-dimensional and of Kac-Moody type. In general, the groups can be thought of as arising as the closure of the area preserving diffeomorphisms $SL(d,\mathbb{R})$ of the M-theory torus and the classical $SO(d-1,d-1,\mathbb{R})$ symmetry realising continuous T-duality~\cite{ObersUDualMTheory}. The possible compactifications are labelled by the classical moduli space $\mathcal{M}^{\textrm{(cl)}}_D= E_{d}(\mathbb{R})/ K(E_{d}(\mathbb{R}))$, where $ K(E_d(\mathbb{R}))$ is the maximal compact subgroup of $E_{d}(\mathbb{R})$.

In string theory,  these continuous symmetries are expected to be broken to the discrete $E_{d}(\mathbb{Z})$ U-duality group, via a Dirac-Schwinger-Zwanziger type quantisation condition related to the existence of charged states (branes)~\cite{HullTownsend,Font:1990gx}. Table~\ref{dualitygroups} shows a complete list of the U-duality groups.
The effect of these discrete dualities is to identify classically inequivalent compactifications: The quantum moduli space of string compactifications to $D$ dimensions is given by ($D=11-d$)
\begin{align}
\label{qms}
\mathcal{M}_D= E_{d}(\mathbb{Z})\backslash E_{d}(\mathbb{R})/ K(E_{d}(\mathbb{R})).
\end{align}

\begin{table}[t]
\begin{center}
\begin{tabular}{ | c || c  c  c  | }
  \hline                       
  $D$ & $E_{d+1}(\mathbb{R})$ & $K(E_{d+1})$ & $E_{d+1}(\mathbb{Z})$ \\ \hline \hline
 $10B$ & $SL(2,\mathbb{R})$ & $SO(2)$ & $SL(2,\mathbb{Z})$ \\ \hline
 $9$ & $\mathbb{R}^+\times SL(2,\mathbb{R})$ & $SO(2)$ & $ SL(2,\mathbb{Z})$ \\ \hline
 $8$ & $SL(2,\mathbb{R})\times SL(3,\mathbb{R})$ & $SO(3)\times SO(2)$ & $SL(2,\mathbb{Z})\times SL(3,\mathbb{Z})$ \\ \hline
 $7$ & $SL(5,\mathbb{R})$ & $SO(5)$ & $SL(5,\mathbb{Z})$  \\ \hline
 $6$ & $SO(5,5,\mathbb{R})$ & $SO(5) \times SO(5)$ & $SO(5,5,\mathbb{Z})$ \\ \hline
 $5$ & $E_{6}(\mathbb{R})$ & $USp(8)$ & $E_{6}(\mathbb{Z})$  \\ \hline
 $4$ & $E_{7}(\mathbb{R})$ & $SU(8)/\mathbb{Z}_2$ & $E_{7}(\mathbb{Z})$ \\ \hline
 $3$ & $E_{8}(\mathbb{R})$ & $Spin(16)/\mathbb{Z}_2$ & $E_{8}(\mathbb{Z})$ \\ \hline
 $2$ & $E_{9}(\mathbb{R})$ & $K(E_{9}(\mathbb{R}))$ & $E_{9}(\mathbb{Z})$ \\ \hline
 $1$ & $E_{10}(\mathbb{R})$ & $K(E_{10}(\mathbb{R}))$ & $E_{10}(\mathbb{Z})$ \\ \hline
  $0$ & $E_{11}(\mathbb{R})$ & $K(E_{11}(\mathbb{R}))$ & $E_{11}(\mathbb{Z})$ \\ \hline
\end{tabular}
\caption{\label{dualitygroups}\em List of the split real forms of the hidden symmetry groups $E_{d(d)}(\mathbb{R})$. We also list the corresponding maximal compact subgroups $K$ and the last column contains the discrete U-duality versions that appear in string theory. The label $10B$ indicates that we are considering type IIB in ten dimensions rather than type~IIA.}
\end{center}
\end{table}

A much studied example where U-duality is explicitly manifest is the type IIB superstring scattering amplitude of the four-graviton scattering process, see e.g.~\cite{GutperleGreen,Green:2010wi,GreenESeries}. More precisely, the amplitude of this process in $D$ dimensions displays an invariance under the respective U-duality group discussed above. Instead of looking directly at the amplitude, one may also consider the corresponding low-energy effective action, where it is found that there is an infinite number of higher-order curvature corrections beyond the Einstein-Hilbert term, of the form
\begin{align}
\label{kcor}
(\alpha')^\frac{2-D}{2}\sum_k \int \mathrm{d}^Dx\,(\alpha')^{k+3}\mathcal{E}^D_{(p,q)}(\Phi)\partial^{2k}R^4\,.
\end{align}
These corrections constitute an expansion in orders of the Regge slope $\alpha'$ (of dimension (length)$^2$) and $R^4$ is given by a specific contraction of four Riemann tensors~\cite{Gross:1986iv}. The first few terms in this expansion beyond the Einstein-Hilbert term occur for $k=2p+3q=0,2,3,4,\ldots$.  The couplings $\mathcal{E}^D_{(p,q)}(\Phi)$ of these terms are functions of the moduli $\Phi\in\mathcal{M}_D$ of the classical moduli space~(\ref{qms}). 

The preservation of U-duality by the corrections in (\ref{kcor}) puts strong constraints on the functions $\mathcal{E}^D_{(p,q)}$. Since the graviton is invariant under U-duality (in Einstein frame), each function $\mathcal{E}^D_{(p,q)}$ has to be a function on $E_{d}(\mathbb{R})/K(E_{d}(\mathbb{R}))$ that is invariant under $E_{d}(\mathbb{Z})$. Furthermore, consistency with string perturbation theory requires that  $\mathcal{E}^D_{(p,q)}$ must have a well-defined `weak coupling' expansion near the weak coupling cusp on $\mathcal{M}_D$. Finally, supersymmetry imposes differential equations on $\mathcal{E}^D_{(p,q)}$~\cite{Green:1998by,Green:2010wi}; for the lowest order corrections $k=0,2$ these differential equations are homogeneous Laplace eigenvalue equations. Altogether, this means that  $\mathcal{E}^D_{(p,q)}$ should be an $E_d(\mathbb{Z})$ automorphic function on $\mathcal{M}_D$. 

In some cases, these constraints are actually strong enough to identify  $\mathcal{E}^D_{(p,q)}$ uniquely~\cite{GutperleGreen,Pioline:1998mn,Green:2010wi} as a non-holomorphic Eisenstein series and there is good evidence that for $k=0,2$ (corresponding to the so-called 1/2-BPS and 1/4-BPS couplings $R^4$ and $\partial^4R^4$) the solution in any dimension $D\geq 3$ is given by such an Eisenstein series~\cite{Green:2010wi,Pioline:2010kb,GreenESeries,GreenSmallRep}. For even higher derivative terms with $k>2$, the Laplace equation becomes inhomogeneous and the automorphic function that is required is unlikely to be an Eisenstein series in general. 

\begin{figure}[t]
\centering
\begin{tikzpicture}
[place/.style={circle,draw=black,fill=black,
inner sep=0pt,minimum size=6}]
\draw (0,0) -- (1,0);
\draw (1,0) -- (1.5,0);
\draw[dashed] (1.5,0) -- (2.5,0);
\draw (2.5,0) -- (3,0);
\draw (3,0) -- (3,1);
\draw (3,0) -- (4,0);
\draw (4,0) -- (5,0);
\node at (0,0) [place,label=below:$\alpha_{d}$] {};
\node at (1,0) [place,label=below:$\alpha_{d-1}$] {};
\node at (3,0) [place,label=below:$\alpha_4$] {};
\node at (4,0) [place,label=below:$\alpha_3$] {};
\node at (5,0) [place,label=below:$\alpha_1$] {};
\node at (3,1) [place,label=above:$\alpha_2$] {};
\end{tikzpicture} 
\caption{\em Dynkin diagram for $E_{d}$. \label{fig:Edplus1Diag}}
\end{figure}
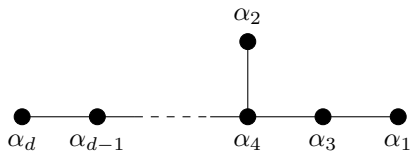

The purpose of the paper~\cite{Fleig:2012xa} ---on which this talk is based--- was to extend the analysis of~\cite{GreenESeries} to $D<3$ for the cases $k=0,2$. This involves generalising the notion of Eisenstein series to Kac-Moody groups since for $D<3$ the hidden symmetries $E_{11-D}$ are of Kac-Moody type, see Table~\ref{dualitygroups}. Pioneering work for Eisenstein series over loop groups was carried out by Garland~\cite{GarlLoop}. We find surprising simplifications for these and more general Eisenstein series as will be shown below. The Dynkin diagram of $E_d$ with our labelling conventions is shown in Figure~\ref{fig:Edplus1Diag}.

These notes are structured as follows. In section~\ref{defn} we will first discuss a general definition of an Eisenstein series, which applies both to the finite- and infinite-dimensional groups. In order to make this abstract definition more transparent, we will derive from it the explicit form of the series over $SL(2,\mathbb{R})$.
In section~\ref{expan} we discuss Fourier expansions of Eisenstein series and provide Langlands' formula for the zero-mode terms of such an expansion, which again is valid for the finite and more general Kac-Moody groups. In the same section we will show that particular Eisenstein series exhibit drastic simplifications in the structure of the constant terms. We will argue that these are the cases that are relevant in string theory and review some consistency checks on our claims.
Finally we provide a short outlook.  More details on many of the points discussed here can be found in~\cite{Fleig:2012xa}.

\section{Eisenstein Series on Kac-Moody groups}\label{defn}

Eisenstein series are functions defined on a non-compact, semisimple real Lie group $G$ and display invariance under a discrete subgroup $G(\mathbb{Z})$ of $G$, see for example~\cite{Langlands}. The invariance property is achieved by defining the series as a sum over orbits of $G(\mathbb{Z})$, typically quotienting by the stabiliser of a cusp to avoid overcounting. 

To make this more transparent, consider the (non-holomorphic) $SL(2,\mathbb{R})$ Eisenstein series invariant under the discrete group $SL(2,\mathbb{Z})$. This series is normally defined as a sum over integers $c$ and $d$ which are co-prime
\begin{align}\label{SL2}
E^{SL(2,\mathbb{Z})}_{s}(\tau)=\sum_{(c,d)=1}\frac{\tau_2^s}{|c\tau+d|^{2s}}\,.
\end{align}
Here $s$ is a complex parameter and the sum is restricted, such that $(c,d)\neq (0,0)$. The argument of the series is $\tau=\tau_1+i\tau_2=\chi+ie^{-\phi}$, which lives in the upper-half of the complex plane. The variables $\phi$ and $\chi$ parameterise the classical moduli $\mathcal{M}_{10}^{\textrm{(cl)}}$ of uncompactified type IIB string theory and are identified as the dilaton and the axion field. 
The function (\ref{SL2}) is clearly non-holomorphic due the appearance of the modulus in the denominator. The group $SL(2,\mathbb{Z})$ acts in the standard fractional linear fashion on $\tau$
\begin{align}
\tau \mapsto \frac{a\tau+b}{c\tau+d}\quad\textrm{for}\quad
\begin{pmatrix}
a&b\\c&d
\end{pmatrix}
\in SL(2,\mathbb{Z}).
\end{align}

In order to make the definition of the Eisenstein series over $SL(2)$ given by~(\ref{SL2}) more easily generalizable, we now write it as a sum over $SL(2,\mathbb{Z})$ orbits. This is possible by realising that the summand in~(\ref{SL2}) can be written as
\begin{align}
\frac{\tau_2^s}{|c\tau+d|^{2s}}= \left[\textrm{Im}(\gamma\cdot \tau)\right]^s \quad\textrm{for}\quad 
\gamma=\begin{pmatrix}a&b\\c&d\end{pmatrix}
\in SL(2,\mathbb{Z}).
\end{align}
The matrix $\gamma$ in this equation is not uniquely defined by $c$ and $d$. But by invoking the modularity condition $ad-bc=1$ all possible solutions for $a$ and $b$ can be obtained from a particular solution $(a_0,b_0)$ through
\begin{align}
\begin{pmatrix} a & b \\
c & d \end{pmatrix}
=
\begin{pmatrix} 1 & m \\
0 & 1 \end{pmatrix}
\begin{pmatrix}a_0 & b_0 \\
c & d \end{pmatrix}
=\begin{pmatrix} a_0+mc & b_0+md \\
c & d \end{pmatrix}
\end{align}
with $m\in\mathbb{Z}$. The shift matrices form the Borel subgroup $B(\mathbb{Z})$ of $SL(2,\mathbb{Z})$:
\begin{align}
B(\mathbb{Z}) = \left\{ \begin{pmatrix}1&m\\0&1\end{pmatrix}: m\in\mathbb{Z}\right\}
\end{align}
and these matrices leave $\textrm{Im}(\tau)=\tau_2$ invariant. Therefore the Eisenstein series (\ref{SL2}) can be written equivalently as a sum over a coset
\begin{align}
E^{SL(2,\mathbb{Z})}_{s}(\tau)=\sum_{\gamma\in B(\mathbb{Z})\backslash SL(2,\mathbb{Z})}\left[\text{Im}(\gamma\cdot \tau)\right]^s.
\end{align}
Finally, we notice that $\textrm{Im}(\tau)$ corresponds to the projection onto the abelian torus of a group element $g\in SL(2,\mathbb{R})$ written in Iwasawa decomposition
\begin{align}
g=nak= \left(\begin{array}{cc} 1 & \tau_1 \\
				   0 & 1 \end{array} \right)
				    \left(\begin{array}{cc} \tau_2^{1/2} & 0 \\
				   0 & \tau_2^{-1/2} \end{array} \right)k
\end{align}
with $k\in SO(2,\mathbb{R})$. Defining the projection to the Cartan subalgebra
\begin{align}
H(g) = \log(a) = \frac12 \log(\tau_2) \cdot h_1, \quad \textrm{where}\quad 
h_1=\begin{pmatrix}
1&0\\0&-1\end{pmatrix}
\end{align}
is the standard $SL(2,\mathbb{R})$ Cartan generator, we see that as another equivalent form of (\ref{SL2}) one obtains
\begin{align}
\label{SL2Lang}
E^{SL(2,\mathbb{Z})}_{s}(\tau)=E^{SL(2,\mathbb{Z})}(\lambda,g)=\sum_{\gamma \in B(\mathbb{Z})\backslash SL(2,\mathbb{Z})}e^{\langle\lambda+\rho|H(\gamma g)\rangle}\,,
\end{align}
where $\lambda=2s \Lambda_1 -\rho$ if $\Lambda_1$ denotes the unique fundamental weight of $\mathfrak{sl}(2,\mathbb{R})$ and $\rho$ the Weyl vector (equal to $\Lambda_1$ here). The angled brackets represent the action of a weight on the Cartan subalgebra element, using $\langle\Lambda_1|h_1\rangle=1$. The addition and subtraction of the Weyl vector might seem a bit awkward here but it turns out that this is convenient for the general theory. 

Expression (\ref{SL2Lang}) is the form of the non-holomorphic Eisenstein series that lends itself to a straightforward generalisation to other groups. This generalised definition for groups of arbitrary rank is
\begin{align}\label{ESeriesFin}
E^G(\lambda,g)\equiv\sum_{\gamma \in B(\mathbb{Z})\backslash G(\mathbb{Z})}e^{\langle\lambda+\rho|H(\gamma g)\rangle}\,.
\end{align}
and was given for finite-dimensional $G$ by Langlands in~\cite{Langlands}. We will only be interested in cases when the weight $\lambda$ appearing in the definition is given by $\lambda = 2s \Lambda_{i_*} -\rho$ 
with $\Lambda_{i_*}$ the fundamental weight of node $i_*$ of the Dynkin diagram of $G$. In that case we denote the Eisenstein series by
\begin{align}
\label{paraweight}
E^G_{i_*;s} (g) \equiv E^G(\lambda,g)\quad \textrm{for}\quad \lambda = 2s \Lambda_{i_*} -\rho.
\end{align}
Often we will leave out the argument $g\in G$ as well. The function $E^G_{i_*;s}$ will be referred to as a maximal parabolic Eisenstein series.

We note that the Eisenstein series of~(\ref{ESeriesFin}), satisfies the Laplace eigenvalue equation
\begin{align}\label{Laplace}
\Delta^{G/K} E^G(\lambda,g) = \frac12\left( \langle \lambda|\lambda\rangle -\langle \rho|\rho\rangle\right) E^G(\lambda,g)\,,
\end{align}
where $\Delta^{G/K}$ is a Laplacian defined on the fields which parameterise the coset $G/K$.\\

The definition of the Eisenstein series as given in~(\ref{ESeriesFin}) also applies in the case when $G$ is a general Kac-Moody group and works in particular for affine $E_9$~\cite{GarlLoop}, the hyperbolic Kac-Moody group $E_{10}$ and the group $E_{11}$~\cite{Fleig:2012xa}. In $D=2$ space-time dimensions the situation is a bit special, since the corresponding U-duality group $E_9$ is an affine Kac-Moody group. The algebra of such a (non-twisted) affine group is constructed from the algebra of the underlying finite-dimensional algebra $\mathfrak{g}$ as
\begin{align}
\hat{\mathfrak{g}} = \mathfrak{g}[[t,t^{-1}]]\oplus c \mathbb{R}\oplus d\mathbb{R}\,,
\end{align}
where the first summand represents the loop algebra of $\mathfrak{g}$, the second summand is associated with the central element and the last summand is the derivation which is counting the affine level. The corresponding Cartan subalgebra $\hat{\mathfrak{a}}$ has dimension $\textrm{dim}(\mathfrak{a})+2$. The definition of an Eisenstein series over affine groups has been worked out rigorously by Garland in~\cite{GarlLoop} and convergence of the series was proven for sufficiently large real parts of the weight defining the Eisenstein series. The definition of the affine Eisenstein series differs subtly from the one of~(\ref{ESeriesFin}), in that one has to include a parameter $v$ in the exponential, which parameterises the group associated with the derivation $d$. For the purpose of the presentation here, we will largely ignore this special case and refer the reader to~\cite{Fleig:2012xa} where its details are treated.

Let us also mention that Eisenstein over infinite-dimensional groups similarly satisfy the eigenvalue equation~(\ref{Laplace}). Subtleties arise again for the case of $E_9$, which are linked to the appearance of scale invariance of gravity in $D=2$ space-time dimensions~\cite{Fleig:2012xa}.

We are now in the position to state the proposed automorphic functions  $\mathcal{E}^D_{(p,q)}$ that appear in the four-graviton scattering process as discussed in the introduction. For the terms $R^4$ and $\partial^4 R^4$ and $D\ge 3$ these functions are given by~\cite{GreenESeries,Obers:1999um,Pioline:2010kb}
\begin{align}\label{ESeries}
\mathcal{E}^D_{(0,0)}=2\zeta(3)E_{1;3/2}^G,\quad\text{and}\quad\mathcal{E}^D_{(1,0)}=\zeta(5)E_{1;5/2}^G\,.
\end{align}
This proposal has passed a variety of checks in the references just given. In~\cite{Fleig:2012xa} we propose that these expressions are also correct for $D<3$ when the symmetry group $G$ becomes infinite-dimensional. In $D=2$, the proposal has to be modified slightly to accommodate properly the derivation $d$~\cite{Fleig:2012xa}.

\section{Fourier expansions of Eisenstein series}\label{expan}

The physical information of the automorphic functions $\mathcal{E}^D_{(p,q)}$ is encoded in their Fourier expansion. In the present work we are only interested in the zero-mode Fourier terms of the expansion. Mathematically, these correspond to the constant terms of the automorphic functions; physically, they represent the perturbative contributions to the scattering process. Although  referred to as the constant term the expressions do depend on the Cartan subalgebra degrees of freedom contained in $A$ of the Iwasawa decomposition of $G=NAK$. The constant term is obtained by integrating out the degrees of freedom contained in the unipotent radical $N$. There exists a formula for the constant term due to Langlands~\cite{Langlands} given by
\begin{align}\label{minparabexp}
\int\limits_{N(\mathbb{Z})\backslash N(\mathbb{R})} E^G(\lambda,g) \mathrm{d}n = \sum_{w\in\mathcal{W}} M(w,\lambda) e^{\langle w\lambda + \rho | H(g)\rangle}\,,
\end{align}
where the sum over the Weyl group $\mathcal{W}$ of $G$ is due to the Bruhat decomposition of $G$. The factor $M(w,\lambda)$ is defined as
\begin{align}\label{M}
M(w,\lambda)=\prod_{\substack{\alpha \in \Delta_+\\w\alpha \in \Delta_-}}\frac{\xi\left(\langle\lambda|\alpha\rangle\right)}{\xi \left(1+\langle\lambda|\alpha\rangle\right)}
=\prod_{\substack{\alpha \in \Delta_+\\w\alpha \in \Delta_-}} c\left(\langle \lambda|\alpha\rangle\right)\,.
\end{align}
The function $\xi$ is the completed Riemann zeta function and its relation with the Riemann $\zeta$ function is $\xi(k)\equiv\pi^{-k/2}\Gamma\left(\frac{k}{2}\right)\zeta(k)$. The sets $\Delta_\pm$ represent the positive/negative roots of the Lie algebra of $G$.  We will analyse the structure of the factor $M(w,\lambda)$ in some more detail in a moment. In particular we will see that its properties are responsible for drastic simplifications in the constant term of the Eisenstein series of~(\ref{ESeries}). We refer to the type of expansion given in~(\ref{minparabexp}) as a minimal parabolic expansion of the constant term.

As an example, let us consider the $SL(2,\mathbb{Z})$ Eisenstein series (\ref{SL2Lang}). The Weyl group has two elements and the application of the constant term formula (\ref{minparabexp}) gives
\begin{align}
\int_{N(\mathbb{Z})\backslash N(\mathbb{R})} E^{SL(2,\mathbb{Z})}_s(g) \mathrm{d}n= \int_0^1 \mathrm{d}\tau_1 E^{SL(2,\mathbb{Z})}_s(\tau) = \tau_2^s + \frac{\xi(s)}{\xi(s+1)} \tau_2^{1-s}.
\end{align}
The two terms have a very precise interpretation from string scattering calculations. The first term corresponds to the string tree level contribution and the second one to the string one-loop result~\cite{Green:1981ya}. There are no further perturbative corrections beyond one-loop due to supersymmetry and the numerical coefficients of string theory agree perfectly with those of the Eisenstein series~\cite{GutperleGreen}.\\

There is also a second type of expansion which is possible that is commonly referred to as a maximal parabolic expansion, where one integrates out the degrees of freedom of the unipotent factor $N_{j_\circ}$ in a particular maximal parabolic subgroup $P_{j_\circ}=N_{j_\circ}M_{j_\circ}$. Here, $j_\circ$ labels a choice of a simple root, with respect to which the maximal parabolic subgroup is defined~\cite{GreenESeries,Fleig:2012xa}. The factor $M_{j_\circ}$ is called the Levi factor in the decomposition of $P_{j_\circ}$. The Levi factor itself can be written as
\begin{align}\label{Levi}
M_{j_\circ}=GL(1)\times G_{d-1}\,,
\end{align}
where $G_{d-1}$ is the group with the Dynkin diagram that is left after deleting the $j_\circ$th node from the diagram of $E_{d}$. The $GL(1)$ factor in this product is parameterised by a single scalar $r\in\mathbb{R}^\times$. Langlands' formula for the constant term in a maximal parabolic expansion then becomes~\cite{Moeglin}
\begin{align}\label{maxparabexp}
\int\limits_{N_{P_{j_\circ}}(\mathbb{Z})\backslash N_{P_{j_\circ}}(\mathbb{R})} 
\!\!\!\!\!\!\!\! E^G(\lambda,g) \mathrm{d}n = \sum_{w\in \mathcal{W}_{j_\circ}\backslash\mathcal{W}} M(w,\lambda)e^{\langle\left(w\lambda+\rho\right)_{\parallel j_\circ}|H(g)\rangle}E^{G_d}\left(\left(w\lambda\right)_{\perp j_\circ},g\right)\,.
\end{align}
Here the notation $(\lambda)_{\parallel j_\circ}$ denotes a projection operator on the component of $\lambda$ which is proportional to the fundamental weight $\Lambda_{j_\circ}$ and $(\lambda)_{\perp j_\circ}$ is orthogonal to $\Lambda_{j_\circ}$.

As for the definition of the Eisenstein series, Langlands' formul\ae~above also apply in the case of the infinite-dimensional groups $E_{10}$ and $E_{11}$. For the affine case slight modifications have to be made again, in order to account for the derivation $d$~\cite{GarlLoop}. Let us also mention that in this case, the Levi factor $M_{j_\circ}=GL(1)\times GL(1)\times G_{d-1}$. There are now two $GL(1)$ factors instead of only one as in~(\ref{Levi}). One factor corresponds to the central element $c$ and the other to the derivation $d$. Hence we now have an additional parameter $v\in\mathbb{R}^\times$ appearing in the expression for the constant term besides $r$.\\

It is easy to see from~(\ref{minparabexp}) that the number of terms that make up the constant term is bounded from above by the order of the Weyl group $\mathcal{W}$. In particular, in the case of finite-dimensional groups, where the order of the Weyl group is also finite, this number is always finite and generically equal to the order of the Weyl group. For the particular choices $s=3/2$ and $s=5/2$ in~(\ref{ESeries}), however, the number reduces drastically~\cite{GreenESeries,Pioline:2010kb} in such a way that only very few non-zero terms are left. This is due to the structure of the coefficient $M(w,\lambda)$ of~(\ref{M}) and physically related to the BPS-ness of the $R^4$ and $\partial^4 R^4$ terms as was studied in detail for $D\ge 3$ in~\cite{GreenESeries} and related to minimal and next-to-minimal automorphic representations in~\cite{GRS,Pioline:2010kb,GreenSmallRep}. 

When considering the infinite-dimensional symmetry groups $E_d(\mathbb{R})$ for $d>8$ ($D<3$) the situation is much less clear. An application of the formula~(\ref{minparabexp}) would lead generically to an infinite number of constant terms since the order of the Weyl group for indefinite Kac-Moody algebras is infinite. The remarkable result of our work~\cite{Fleig:2012xa} was that for the special values $s=3/2$ and $s=5/2$ this generic number reduces to a finite number as required by physical arguments.

Let us now explain the mechanism for this simplification along with a practical implementation~\cite{Fleig:2012xa}, see also~\cite{GreenESeries}. The function $M(w,\lambda)$ of~(\ref{M}) is of central importance. It satisfies the multiplicative property
\begin{align}
\label{cocycle}
M(w\tilde{w},\lambda) = M(w,\tilde{w}(\lambda)) M(\tilde{w},\lambda).
\end{align}
The function $c(k)$ that appears in the factors that contribute to $M(w,\lambda)$ have special values only at arguments $k=\pm1$, namely
\begin{align}
c(-1) = 0\,,\quad\quad c(+1) = \infty\,
\end{align}
and $c(-1)c(1)=1$. This means that if, for a particular Weyl word $w\in\mathcal{W}$, the product giving $M(w,\lambda)$ contains more $c(-1)$ than $c(1)$ factors, then $M(w,\lambda)$ will be zero. In addition, one can show by the multiplicative property (\ref{cocycle}) that if $M(\tilde{w},\lambda)=0$ for some $\tilde{w}$, then any longer Weyl word of the form $w\tilde{w}$ will also lead to $M(w\tilde{w},\lambda)=0$. Moreover, it is easy to see that any $w$ that stabilises $\lambda+\rho$ will lead to $M(w,\lambda)=0$. Restricting to the particular case $\lambda=2s\Lambda_{i_*}-\rho$, the sum over Weyl words reduces therefore at least to the subset
\begin{align}
\label{Si}
\mathcal{S}_{i_*}\equiv\left\{w\in\mathcal{W}|w\alpha_i>0\text{  for all  } i\neq i_*\right\}= \mathcal{W}/\mathcal{W}_{i_*},
\end{align}
where $\mathcal{W}_{i_*}$ is the stabiliser of $\Lambda_{i_*}$. The set $\mathcal{S}_{i_*}$ is in bijection with the Weyl orbit of $\Lambda_{i_*}$ and this also gives a convenient way of enumerating the set in a partially ordered manner. This was shown in~\cite{Fleig:2012xa}. Let us emphasise again that the number of Weyl words in $\mathcal{S}_{i_*}$ is only a `small fraction' of the order of the whole Weyl group $\mathcal{W}$. Of course, this number is still infinite in the Kac-Moody case.

Representing $\mathcal{S}_{i_*}$ as a partially ordered set corresponding to the Weyl orbit of $\Lambda_{i_*}$ also allows us to exploit the full power of~(\ref{cocycle}). We can picture the partially ordered set of Weyl words as a tree rooted at the identity Weyl word. By parsing through the partially ordered set of Weyl words and computing $M(w,\lambda)$ along all branches of the tree, we know by~(\ref{cocycle}) that we can terminate the investigation of a given branch if we reach a vertex $\tilde{w}$ of the tree where $M(\tilde{w},\lambda)$ vanishes. 
In order to determine whether this happens we analyse the factors that contribute to the product~(\ref{M}). For this purpose, we define two different sets of roots $\Delta_s(\pm1)$; one set for all positive roots producing $c(-1)$ factors and the other for roots producing $c(+1)$ factors in the product
\begin{align}
\Delta_s(\pm1) := \left\{ \alpha\textrm{ contributing to $M(w,\lambda)$} \;:\; \langle \lambda|\alpha\rangle 
=\langle 2s \Lambda_{i_*}-\rho | \alpha\rangle =\pm 1\right\}.
\end{align}
These sets are well-defined as there are only finitely many $\alpha$ contributing to $M(w,\lambda)$ for a given $w\in\mathcal{W}$.
By working out how many roots from each of the two sets will contribute in the product defining $M(w,\lambda)$, one can see that for specific choices of $s$ and $i_*$, only a finite number of Weyl words in $\mathcal{S}_{i_*}$ will yield a non-zero $M(w,\lambda)$ factor. It turns out that such a specific choice is given by $s=3/2$ and $5/2$ and $i_*=1$. Therefore to summarise again, we can say that for special choices of maximal parabolic Eisenstein series the constant term, which was na\"ively thought to contain an infinite number of terms, collapses to a finite sum of only a few terms.\\

The next step is to investigate the space of possible values of $s$ across dimensions and in particular for $D<3$. For this we have computed the number of terms in the constant term of a minimal parabolic expansion. Table~\ref{svalues} shows the result for $D\leq5$ for a range of integer and half-integer values for $s$.
\begin{table}
\begin{center} 
\scalebox{.88}{
\begin{tabular}{ | c | c  c  c  c  c  c  c  c  c  c  c  c  c  c | }
 \hline &&&&&&&&&&&&&& \\[-3mm]                      
  $s$ & $0$ & $\frac12$ & $1$ & $\frac32$ & $2$ & $\frac52$ & $3$ & $\frac72$ & $4$ & $\frac92$ & $5$ & $\frac{11}{2}$ & $6$ & $\frac{13}{2}$ \\[1mm] \hline\hline
 $E_6$ & $1$ & $2$ & $27$ & $7$ & $12$ & $27$ & $\cdots$ & $$ & $$ & $$ & $$ & $$ & $$ &  \\ \hline
 $E_7$ & $1$ & $2$ & $126$ & $8$ & $14$ & $35$ & $56$ & $126$ & $91$ & $126$ & $\cdots$ & $$ & $$ &  \\ \hline
 $E_8$ & $1$ & $2$ & $2160$ & $9$ & $16$ & $44$ & $72$ & $408$ & $534$ & $1060$ & $1460$ & $1795$ & $2160$ & $\cdots$ \\ \hline
 $E_9$ & $1$ & $2$ & $\infty$ & $10$ & $18$ & $54$ & $90$ & $\infty$ & $\cdots$ & $$ & $$ & $$ & $$ &  \\ \hline
 $E_{10}$ & $1$ & $2$ & $\infty$ & $11$ & $20$ & $65$ & $110$ & $\infty$ & $\cdots$ & $$ & $$ & $$ & $$ &  \\ \hline
 $E_{11}$ & $1$ & $2$ & $\infty$ & $12$ & $22$ & $77$ & $132$ & $\infty$ & $\cdots$ & $$ & $$ & $$ & $$ &  \\ \hline
\end{tabular}}
\end{center}
\caption{\em The table shows the number of Weyl words with non-vanishing coefficients $M(w,\lambda)$ in an expansion of $E^{E_d}_{1;s}$ in dimensions $0\leq D\leq5 $ and for a range of values for the parameter~$s$. An ellipsis signifies that the row is continued with the last number explicitly written out.
\label{svalues}}
\end{table}
For $D<3$ there seem to be only a few values of $s$, amongst them of course $s=3/2$ and $5/2$, for which one obtains the collapse of the infinite sum explained above. There is however a large number of values for which this collapse does not seem to happen. In particular for values of $s\geq7/2$ the calculation of the constant term on a computer did not terminate within a reasonably short period of time (unlike it did for values of $s<7/2$). This can be taken as a tentative indication that in these cases the number of Weyl words contributing to the sum in Langlands' formula is actually infinite and for this reason we put $\infty$ for the corresponding entries in Table~\ref{svalues}. (Physically, this may be related to counterterms being unprotected by supersymmetry.)

Looking at Table \ref{svalues} it is tempting to interpret it as a strong sign for the special properties associated with the small values of $s$ in the set 
\begin{align}
s\in\left\{0,1/2,3/2,2,5/2,3\right\}\,.
\end{align}
More precisely, by requiring the constant term to only encode a \textit{finite} number of perturbative effects as required by supersymmetry, the range of possible values that $s$ can take, gets reduced from a previously infinite set to a finite number of possible values. 
It would certainly be desirable to make these statements more precise and to prove them rigorously. In our paper~\cite{Fleig:2012xa} we not only compute the number of constant terms but also their precise form. Some of them develop logarithmic dependence on the Cartan subalgebra coordinates.

\section{Remarks and outlook}

The maximal parabolic Fourier expansion~(\ref{maxparabexp}) can be used to check the consistency of the automorphic couplings $\mathcal{E}^D_{(p,q)}$ in the low-energy expansion. Namely, the functions~(\ref{ESeries}) are subject to a number of strong consistency requirements~\cite{Green:2010wi,Pioline:2010kb} that arise from the interplay of string theory in various dimensions. The consistency conditions are typically phrased in terms of three (maximal parabolic) limits, corresponding to different combinations of the torus radii (in appropriate units) and the string coupling becoming large. The three standard limits correspond to 
\begin{enumerate}
\item[(i)] decompactification from $D$ to $D+1$ dimensions, where one torus circle becomes large,
\item[(ii)] string perturbation theory, where the $D$-dimensional string coupling is small, and 
\item[(iii)] the M-theory limit, where the whole torus volume becomes large. 
\end{enumerate}
In terms of the $E_{d}$ diagram this means singling out the nodes $d$, $1$ or $2$, respectively. Mathematically, these limits are tantamount to computing the constant terms of the Eisenstein series in different maximal parabolic expansions~(\ref{maxparabexp}). We have performed these consistency checks for our Eisenstein series (\ref{ESeries}) in the Kac-Moody case. The functions~(\ref{ESeries}) satisfy them for all $D<3$. In the affine case, particular care must be taken due to the scale invariance of the two-dimensional gravity system. Na\"ive evaluations of the Laplace equations and decompactification limits will lead to non-sensical answers. A careful discussion of how to remedy this by the proper inclusion of the derivation and the central charge of the affine algebra can be found in~\cite{Fleig:2012xa}.

Among the interesting future directions, we mention the question of the non-zero-mode Fourier coefficients. The abelian ones are expected to be related to instantons as in higher space-time dimensions~\cite{Pioline:2010kb,GreenSmallRep}, however, the study of instantons in low space-time dimensions bears its own subtleties since finite energy solutions are harder to construct than in higher space-time dimensions. This is due to the asymptotic behaviour of the Green functions of the Laplace operator. We anticipate that the study of the Fourier coefficients might shed some light on this question. Not unrelated is the issue of automorphic representations. The collapse of the constant term can be interpreted as resulting from $\mathcal{E}^D_{(p,q)}$ being associated with a small automorphic representation~\cite{GRS,Pioline:2010kb,GreenSmallRep}. Automorphic representations of Kac-Moody groups have not been studied to the best of our knowledge.

Yet another possible application of automorphic functions and of $E_{10}(\mathbb{Z})$ in particular is in the context of arithmetic quantum gravity as defined in~\cite{Ganor:1999ui,Pioline:2003bk,Brown:2004jb,Kleinschmidt:2009cv}. There it was argued that the wavefunction of the universe should be an automorphic function of $E_{10}(\mathbb{Z})$ with zero eigenvalue under the Laplacian. Whether such a function exists and also satisfies the additional boundary conditions is presently not known.

\subsection*{Acknowledgements}
We would like the organisers of QQQ in Tallinn and in particular the session chair D.~Persson for putting together a broad and interesting programme and an enjoyable conference. For~\cite{Fleig:2012xa}, we benefitted from discussions with T. Damour, S. Fredenhagen, M. Green, H.~Nicolai, S. Miller,  B.~Pioline, P.~Vanhove and D. Persson who also made useful comments on the present manuscript.

\bibliography{tallinproc}
\bibliographystyle{amsplain}

\end{document}